\documentclass[a4paper]{jpconf}
\usepackage{graphicx}
   \usepackage{amsmath}
  \usepackage{amssymb}
\usepackage{multirow}
\usepackage{textcomp}         

\begin{document}
\title{Semi-classical approach to $J/\psi$ suppression in high energy heavy-ion collisions}

\author{R Katz and P B Gossiaux}

\address{SUBATECH, CNRS/IN2P3, Universit\'e de Nantes, Ecole des Mines de Nantes,
4 rue Alfred Kastler, 44307 Nantes cedex 3, France}

\ead{katz@subatech.in2p3.fr}

\begin{abstract}
We study the heavy quark/antiquark pair dynamics in strongly-coupled quark gluon plasma. A semi-classical approach, based on the Wigner distribution and Langevin dynamics, is applied to a color screened $c{\bar c}$ pair, in a hydrodynamically cooling fireball, to evaluate the total $J/\psi$ suppression at both RHIC and LHC energies. Although its limitation is observed, this approach results to a $J/\psi$ suppression of around 0.30 at RHIC and 0.25 at LHC.
\end{abstract}

\section{Introduction}

Quarkonia suppression was predicted by Matsui and Satz \cite{MatsuiSatz} as a sign of Quark-Gluon Plasma production in heavy-ion collisions. It has been experimentally observed but some aspects are still poorly understood, e.g. the unexpected ``suppression" of the $J/\psi$ suppression at low $p_T$ as the collision energy increases. This paper aims to describe the $J/\psi$ suppression from a dynamical point of view, as an alternative scheme to stationary sequential suppression, recombination... 

Instead of studying the bound states survival  in a stationary medium, we focus on the $Q{\bar Q}$ pair ($Q$ being a heavy quark) dynamics in a non-stationary QGP until the freeze out. The QGP is considered here as a color screening medium at thermal equilibrium with a homogeneous time dependent temperature. The elliptic flow observations at RHIC and LHC suggest that the quarkonia thermalise partially with the medium. Our $Q{\bar Q}$ pair will therefore undergo a color screened, temperature dependent, self potential and some Langevin forces coming from its direct interaction with the thermal medium. We expect the thermalisation effect to be the answer to the observed suppression of the suppression as it should tend to hold the $Q{\bar Q}$ together. 

Ideally a full quantum formalism would be desirable, however the thermalisation of a quantum state being still an unsolved problem, we base this study on a semi-classical formalism inspired by Young and Shuryak work \cite{Young:2008he}. As a test of reliability, a comparison to pure quantum results is carried out in the case without Langevin dynamics.

%
%
\section{The Semi-classical approach without Langevin dynamics}\label{SCNoTherm} 

\subsection{The Wigner approach}

In usual quantum mechanics, the $Q{\bar Q}$ pair probabilistic information is described by a wavefunction $\Psi$, which evolution is given by the Schr\"odinger equation. Equivalently, the $Q{\bar Q}$ pair can be described as a phase space distribution called the Wigner distribution $F\left(\vec{r},\vec{p},t\right)$, derived from the Wigner transformation of the wavefunction \cite{Liboff}. Its evolution is then given by the Wigner-Moyal equation, which classical limit (equivalent to the Liouville equation) is used within the semi-classical frame. To evaluate the probability (or "$J/\psi$ weight") of the $Q{\bar Q}$ pair to bind as a $J/\psi$ state, one projects its evolved Wigner distribution onto the $J/\psi$ Wigner distribution, computed from the vacuum potential. This semi-classical approach allows us the access to numerical simulations through the practical use of the test particles method. 
\vspace{-2mm}
\subsection{The color screened potentials and temperature scenarios}\label{Pots}
The binding $Q{\bar Q}$ potential in the QGP is a temperature dependent potential that takes into account the presence of color charges in their vicinity. We choose to lead this study with the potential $U(\vec{r},T)$, evaluated by Kaczmarek et al. \cite{Kaczmarek} from lQCD results and re-parameterised by us. It corresponds to the color singlet internal energy, i.e. assuming no energy exchange between the pair and the medium (a redundancy with the stochastic forces is therefore avoided).

\begin{figure}[!h]
 \centerline{
\includegraphics[height=40mm]{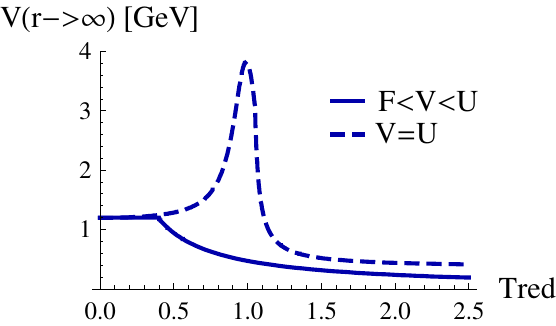}\hspace{10mm}
\includegraphics[height=40mm]{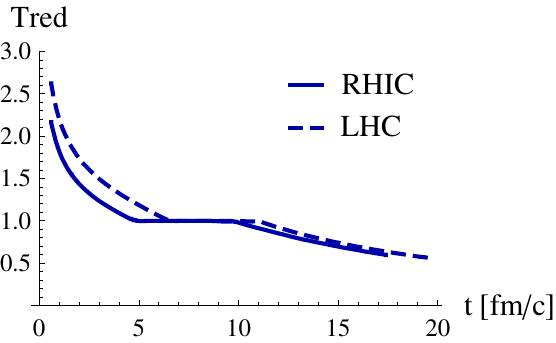}}
   \caption{\label{fig: PotAndTempEvolution}
   \small   {\it Left}: "Intermediate" $F<V<U$ \cite{MocPet1} and internal energy U potential asymptotic dependences on temperature. {\it Right}: Evolution of the reduced temperature $T_{red}(t)=T/T_c$ over time at the centre of the fireball at RHIC ($\sqrt{s_{NN}}=200$ GeV) and LHC ($\sqrt{s_{NN}}=2.76$ TeV) energies.}
\end{figure}

In the common heavy-ion collisions simplified scenario, the $Q{\bar Q}$  pair is first produced within a short lapse of time ($\tau\lesssim 0.3$ fm/c), promptly followed by a warm cooling QGP phase ($T\gtrsim T_c$ and $0.3\lesssim\tau\lesssim 5$ fm/c) in which the $Q{\bar Q}$ pair evolves. Finally, a cooler hadronisation phase ($T^{had}< T_c$) follows until the freeze out. The critical temperature $T_c$ is taken equal to 165 MeV. We choose the space-time temperature distribution derived by \cite{KolbHeinz} through the use of a hydrodynamic evolution of an initial thermalised state \cite{Glauber}. To simplify, the spatial distribution is reduced to the temperature at the centre of the fireball (see figure \ref{fig: PotAndTempEvolution}) and the $Q{\bar Q}$ pair is assumed to be produced at $t=0.6$ fm/c, i.e. from which the QGP is at thermal equilibrium. 
\vspace{-7mm}
\subsection{Initial distribution}\label{Idist}

As we are only interested in S states, the $\Psi_{Q{\bar Q}}({\bf r},t)$ wave-function is reduced to its radial part $R_{Q{\bar Q}}(r,t)$. The initial radial wave-function is a Gaussian wavepacket with a 0.165 fm r.m.s. for the $c{\bar c}$ pair. The latter can be estimated by applying the uncertainty principle to the intermediate quark of the $Q{\bar Q}$ production Feynman diagrams (LO u or t channel): $\Delta r \sim \hbar c/m_c\sim 0.16$ fm. 
\vspace{-2mm}
\subsection{Semi-classical and quantum results without thermalisation processes}\label{ResultsNoTherm} 

\begin{figure}[!h]
 \centerline{
\includegraphics[height=40mm]{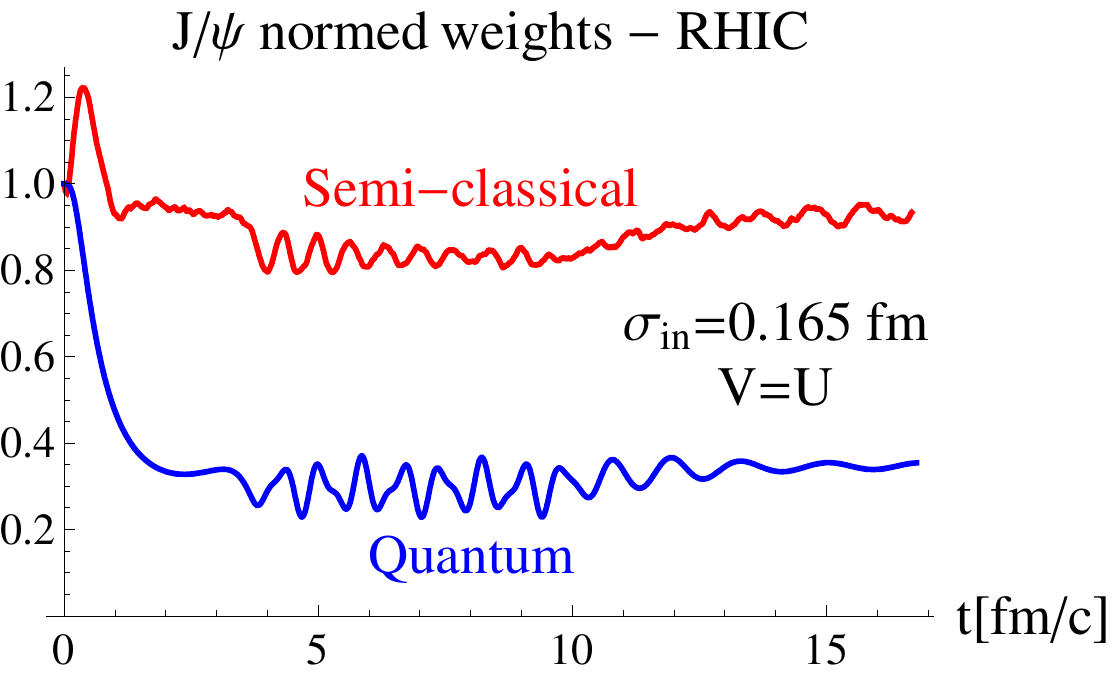}\hspace{10mm}
\includegraphics[height=40mm]{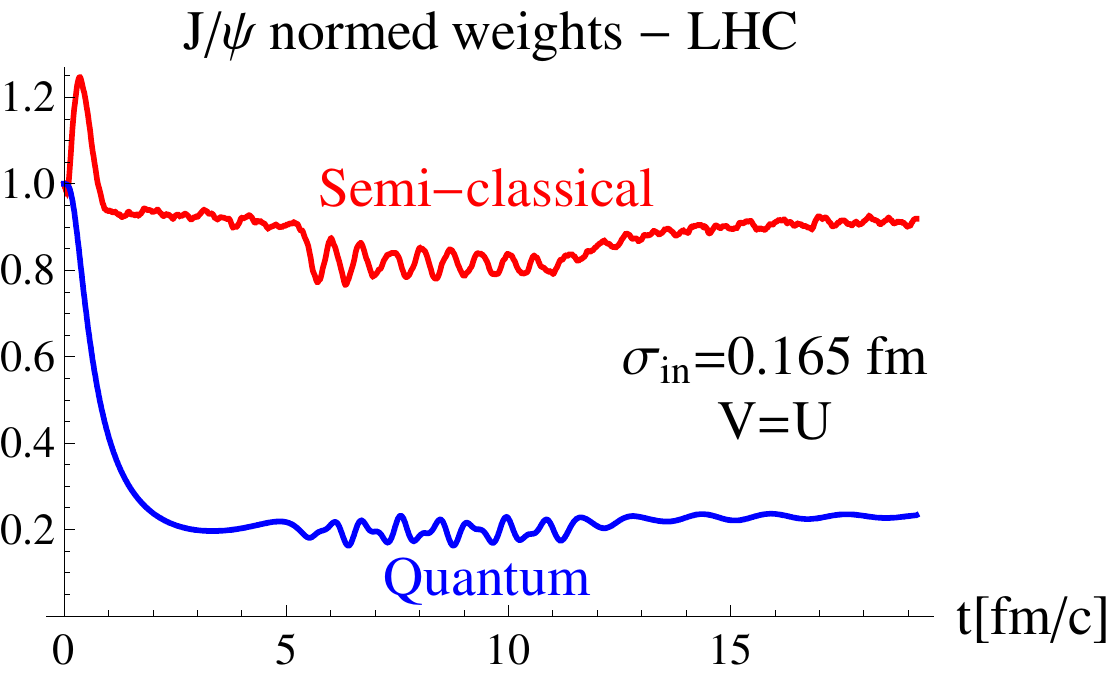}}
  \caption{\label{fig:JpsiWeightsRHICLHC}
   \small   Semi-classical and quantum results for the $J/\psi$ weights function of time with RHIC ({\it left}) and LHC ({\it right}) temperature scenarios.}
\end{figure}

In order to check the limits of the semi-classical approximation, the results are here compared to pure quantum results (from the time-dependent Schr\"odinger equation). If V=0, the $J/\psi$ weights are identical as expected from the Erhenfest theorem. However, in the case of our screened potential, the semi classical results exhibit strong discrepancies (Fig. \ref{fig:JpsiWeightsRHICLHC}): a ``lump" for $t<1$ fm/c and a difficulty to reach the continuum (the $J/\psi$ normed weight remains close to 1). The observation of the test particle paths in phase space shows that the lump is due to their loss of momentum while climbing the potential barrier (making them enter the ``$J/\psi$ zone"). Consequently, if the Langevin dynamics does not lead the evolution, the validity of the semi-classical ``thermalised'' results (section \ref {SCTherm} and \cite{Young:2008he}) is questionable. As a positive point, one observes for both formalism oscillations between eigenstates due to high asymptotic values of the potential when $5 \lesssim t \lesssim 12$ fm/c.

\section{The Semi-classical approach with Langevin dynamics}\label{SCTherm} 

\subsection{Semi-classical formalism with Langevin dynamics}\label{FormalismSCTherm}

A way to take into account the thermalisation of the $Q{\bar Q}$ pair in this dynamical model is to consider the random interactions between the $Q{\bar Q}$ pair and the QGP constituents. By analogy with the Fokker-Planck equation of motion in momentum space (equivalent to Langevin forces), we introduce additional stochastic terms in the Wigner Moyal equation.

In practice (particle test method), Langevin forces are introduced in Newton's equation of motion: a stochastic force $\vec{\xi}$, defined by $\langle\,\vec{\xi}\,\rangle$=0 and $\langle\xi_i (t) \xi_j (t')\rangle$ = $\kappa \delta_{i,j} \delta$(t-t') (the fluctuations are uncorrelated over time), and a "friction" term $-A\,\vec{p}$ where $A$ is the drag coefficient. The Einstein relation can then be deduced from quadratic and average momentum calculations: $\kappa/2= m T/A$. A microscopic calculation from Gossiaux and Aichelin and a fit to experimental $R_{AA}$ lead to a $T$-dependent drag: $A[\mbox{c/fm}] \cong 3 T[\mbox{GeV}] + 2.5 T^{2}$ \cite{Gossiaux:2008jv} for charm quarks. By mean of comparison, we will also use Young and Shuryak's drag $A=4\pi T^2/(3\hbar c m)$ \cite{Young:2008he}.

\vspace{-2mm}
\subsection{Results with RHIC and LHC temperature scenarios}\label{ResultsSCTherm}
\begin{figure}[!h]
 \centerline{
\includegraphics[height=40mm]{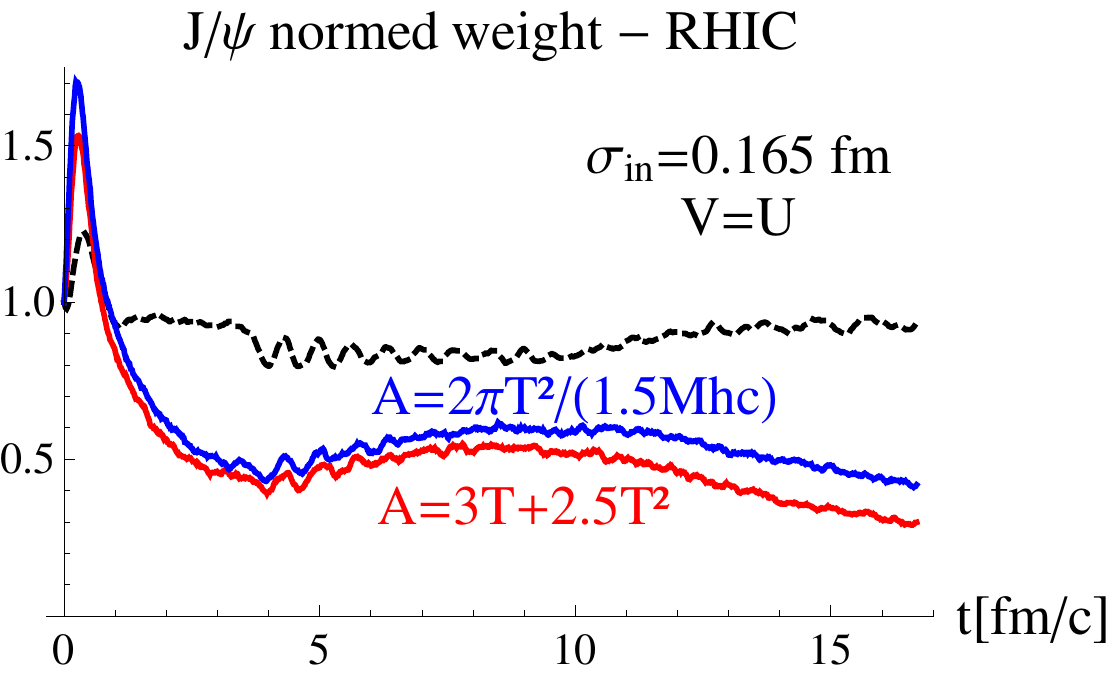}\hspace{10mm}
\includegraphics[height=40mm]{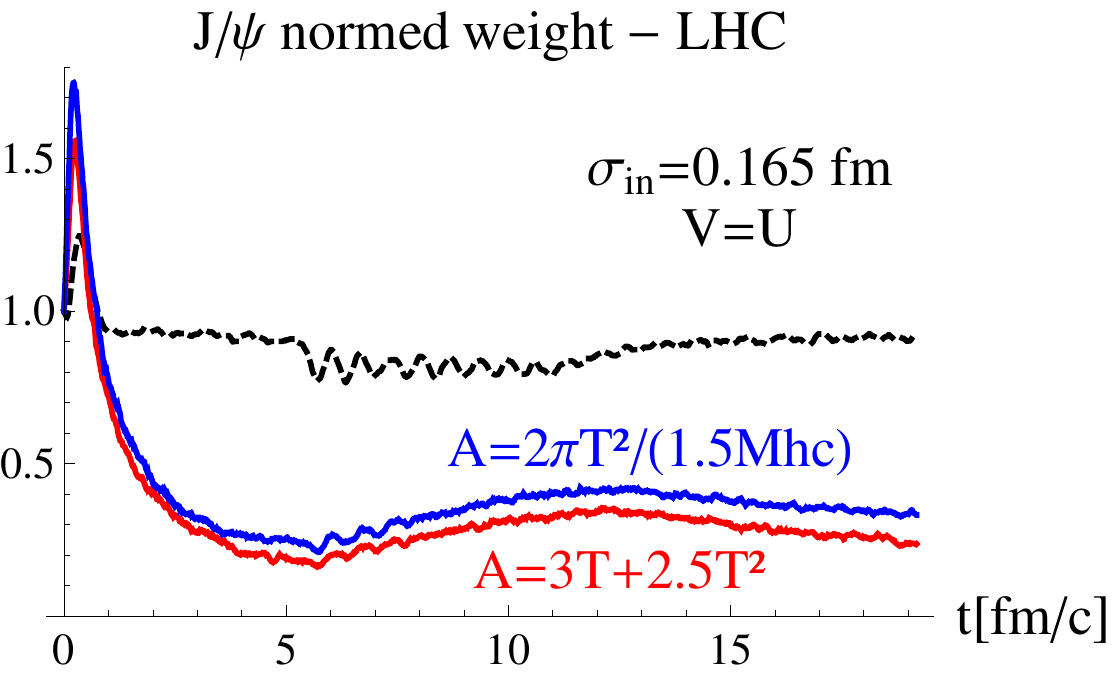}}
  \caption{\label{fig:JpsiWeightsRHICLHCTerm}
   \small  Semi-classical results with stochastic forces for the $J/\psi$ weights function of time at RHIC ({\it left}) and LHC ({\it right}) and the two drags. Dashed lines: with no stochastic forces.}
\end{figure}

As shown in Fig.\ref{fig:JpsiWeightsRHICLHCTerm}, the additional Langevin dynamics leads to an actual evolution of the $J/\psi$ weights that was missing in section \ref{ResultsNoTherm}: not only the $c{\bar c}$ pair thermalises with the QGP, but it also solves the previous test particles difficulty to reach the continuum. It also enhances the ``lump" for $t<1$ fm/c, thanks to rapid thermalisation of the distribution in momentum space \cite{Young:2008he} due to the drag term. The weight slow decrease, for $1 \lesssim t \lesssim 5$ fm/c, comes from the spatial diffusion of the $c{\bar c}$ pair distribution due to the stochastic force term. From $t \gtrsim 5$ fm/c, the light weight variations follow the important variations of the U potential: high asymptotic values lead to a narrowing of the spatial distribution and low asymptotic values to its spatial diffusion. 

The two drag coefficients give similar evolutions with a difference of $\sim 0.1$ at the freeze out. As a robustness test, an initial wavepacket r.m.s. (see section \ref{Idist}) variation of $\pm0.01$ fm/c ($12\%$) leads to a maximum normed $J/\psi$ weight deviation of 0.03 ($8\%$) at the freeze out.

\vspace{-2mm}
\section{Conclusion}\label{Concl} 

In order to better understand the $J/\psi$ suppression at RHIC and LHC, the dynamics of a $c{\bar c}$ pair has been studied through a semi-classical formalism proposed by Young and Shuryak \cite{Young:2008he}. This formalism, i.e. the classical evolution of a $Q{\bar Q}$ quantum Wigner distribution, has been chosen as a convenient way to introduce a classical thermalisation process. The normed $J/\psi$ weights obtained at the freeze out with this formalism are summed up in table \ref{SumUpTable}, and compared to some extent to pure quantum results and experimental data. Of course the comparison to data may not be taken too seriously as we have not considered cold nuclear matter effects, feed downs from other quarkonia... Including them would mostly have a positive effect as our results (especially at RHIC) underestimate the data at high $p_T$, where color-screening effects are expected to be relatively more important \cite{Adamczyk:2012ey}.
\begin{table}[h!]
\begin{center}
    \begin{tabular}{|c|c|c|c|c|c|c|}
    \hline 
    & \multicolumn{2}{ |c| }{\vspace{-0.5mm} Semi-classical} & \multicolumn{2}{ |c| }{Quantum} &  \multicolumn{2}{ |c| }{Experimental $R_{AA}$} \\ \cline{1-7}
    \vspace{-0.5mm}Thermalisation ? $\rightarrow$ & No & Yes & No & Yes & high $p_T$ & low $p_T$\\ \hline
    \vspace{-0.5mm}RHIC & 0.9 & 0.29 & 0.31 & ? & 0.64 $\pm$0.14 & 0.26 $\pm$0.05\\ \hline
    \vspace{-0.5mm}LHC & 0.9 & 0.23 &  0.23 & ? & 0.20 $\pm$0.03 & 0.83 $\pm$0.14\\
    \hline
    \end{tabular}
\caption {\label{SumUpTable} 
\small $J/\psi$ normed weights for RHIC and LHC at the freeze out. --- RHIC AuAu $\sqrt{S_{NN}}=200$ GeV collisions: 1) high $p_T$ STAR data \cite{Adamczyk:2012ey} (inclusive (prompt and non prompt) $J/\psi$, $5<p_T<14$ GeV/c, $|y|<1$ and 0-10\% centrality) and 2) low $p_T$ PHENIX data \cite{Adare:2011yf} (inclusive $J/\psi$, $p_T<5$ GeV/c, $|y|<0.35$ and 0-5\% centrality). The non prompt contribution is estimated to 10-25\% of the inclusive production. --- LHC PbPb $\sqrt{S_{NN}}=2.76$ TeV collisions: 1) high $p_T$ CMS data \cite{Chatrchyan:2012np} (prompt $J/\psi$, $6.5<p_T<30$ GeV/c, $|y|<2.4$ and 0-10\% centrality) (inclusive $J/\psi$: $R_{AA}=0.24 \pm$0.03) and 2) low $p_T$ ALICE data \cite{Maire:2013ad} (inclusive $J/\psi$, $0<p_T<8$ GeV/c, $|y|<0.9$ and 0-10\% centrality).}
\end{center}
\end {table}
\vspace{-5mm}

The comparison between quantum and semi-classical results has revealed the limitation of the latter, and makes questionable the accuracy of the results obtained with the additional Langevin dynamics (Table \ref{SumUpTable} and \cite{Young:2008he}). They could however still be relevant if the stochastic forces appear to be the leading ingredient of the evolution. It seems to be indeed the case within the semi-classical frame, but remains to be checked in a full quantum study. In a near future work, a full dynamical quantum approach including some stochastic forces will be developed.

\section*{Acknowledgments}
We are grateful for support from TOGETHER project R\'egion Pays de la Loire.


\end{document}